\documentclass[%
reprint,
superscriptaddress,
 amsmath,amssymb,
 aps,
floatfix,
]{revtex4-1}
\usepackage{graphicx}
\usepackage{color}
\usepackage{subfigure}
\begin{document}

\title{
Size dependence of dynamic fluctuations in liquid and supercooled water}

\author{Joan Manuel Montes de Oca}
\affiliation{INQUISUR, Departamento de Qu\'{i}mica, Universidad Nacional del Sur (UNS)-CONICET, Avenida Alem 1253, 8000 Bah\'{i}a Blanca, Argentina.}
\author{Sebasti\'an R. Accordino}
\affiliation{INQUISUR, Departamento de Qu\'{i}mica, Universidad Nacional del Sur (UNS)-CONICET, Avenida Alem 1253, 8000 Bah\'{i}a Blanca, Argentina.}
\author{Gustavo A. Appignanesi}
\affiliation{INQUISUR, Departamento de Qu\'{i}mica, Universidad Nacional del Sur (UNS)-CONICET, Avenida Alem 1253, 8000 Bah\'{i}a Blanca, Argentina.}
\author{Philip H. Handle}
\affiliation{Department of Physical Chemistry, University of Innsbruck, Innrain 52c, A-6020 Innsbruck, Austria.}
\author{Francesco Sciortino}
\affiliation{Dipartimento di Fisica,  Sapienza Universita' di Roma, Piazzale A. Moro 5, Roma 00185, Italy.}
\affiliation{CNR-ISC, c/o Sapienza, Piazzale A. Moro
5, Roma 00185, Italy.}

\date{\today}

\begin{abstract}
We study the evolution of dynamic fluctuations averaged over different space
lengths and time scales to characterize spatially
and temporally heterogeneous behavior of  TIP4P/2005 water in liquid and supercooled states. Analysing a million particle simulated system we provide evidence of the existence, upon supercooling, of a significant enhancement of spatially localized dynamic fluctuations stemming from regions of correlated mobile molecules.  We   show that both the magnitude of the departure from the value expected for the system-size dependence of an uncorrelated system  and the molecular size at which such trivial regime is finally recovered clearly increase upon supercooling. This provides a means to estimate an upper limit to the maximum length scale of influence of the regions of correlated mobile molecules.
Notably, such upper limit 
grows two orders of magnitude on cooling, reaching a value corresponding to a few thousand molecules at the lowest investigated temperature. 


\end{abstract}


\maketitle

\section{Introduction}

When we cool a liquid fast enough to prevent crystallization we obtain a supercooled liquid that ultimately transforms into a glass, a solid metastable material with disordered liquid-like
structure ~\cite{angell95,angell00}. But even if from a practical point of view this process has been known for centuries, the comprehension of the molecular expedient by which the liquid falls out of equilibrium remains one of the most interesting  topics in condensed matter physics~\cite{langer14,chandler10,biroli13,ediger12,lubchenko07,cavagna09,charbonneau17}. A major breakthrough  was the discovery of  dynamical heterogeneities,
regions of atoms or molecules moving in a cooperative way,  in a spatially and temporally heterogeneous fashion
\cite{sillescu99,ediger00,glotzer00,hempel00,richert02}.
At any particular time certain regions of the sample are virtually frozen
while others are quite mobile and 
characterized by a ``cooperative'' motion where localized groups
of molecules exhibit significant displacements
\cite{glotzer00,kob97,donati98,donati99}.  While early studies~\cite{kob97,donati98,donati99}
used various somewhat arbitrary criteria to define mobile
particles, later work examined spatial correlation functions
averaged over all particles in various ways attempting to
identify the length and time scales of dynamical heterogeneity
\cite{donati99,flenner07,doliwa00,weeks07cor,glotzer00,lacevic03,keys07,appignanesi06, appignanesi09,appignanesi06b, appignanesi07, appignanesi11, toninelli05}.
Particularly useful insights have resulted from four-point correlation functions like the four-point dynamical susceptibility, $\chi_4$, function (see Ref.~\cite{review_chi4} for a comprehensive review).

Slow dynamics in liquid water has also received a significant attention. Water is central for main fields ranging from biology to materials science~\cite{ball_matrix,chandler,berne,giovambattista,italianos,karplus,montgomery,stanley,water_m1,water_m2,
water_m3,water_m4, water_m6,water_m7}. Within such contexts, being usually at interfaces or subject to nanoconfinement, water usually shows certain reminiscences of glassy behavior at low or even at room temperature~\cite{italianos,karplus,montgomery,stanley,water_m1,water_m2,
water_m3,water_m4}. Indeed, pure supercooled water represents a system of huge interest in itself since it exhibits an unusual
behavior whose comprehension still remains incomplete despite intense experimental and
theoretical work on both thermodynamical and dynamical grounds~\cite{debenedetti_96,mishima_98,angell_02,angell_04,
mishima_98,angell_02,angell_04, mishima_98,shiratani_96,shiratani_98,sasai1,sasai2,3,3.1,3.2,waterepje,waterepje2,water_m5,water_m8,water_m9, gallo16-crev,handle2017supercooled,handle12-prl,amann-winkel13-pnas,perakis2017diffusive,handle2018experimental,handle2018potential,review}. Computationally it has been shown that
dynamical heterogeneities are also observed in 
 supercooled water, with mobile molecules arranged in clusters that perform collective relaxing motions ~\cite{la2004static, water_PRE_F}.

Recently, some of us have introduced a new approach
to characterize spatial and temporal dynamical heterogeneity that
does not require any {\it a priori} definition of particle mobility.
This has been achieved by using a parameter-free method that contrasts spatial and temporal motion within
regions of a system with corresponding  quantities 
evaluated in 
the large system limit and averaged over space and time~\cite{PRE-RapidComm}. Specifically, we used the
system average mean square displacement as a ``null hypothesis''
for particle motion and we quantified deviations away from this null
hypothesis by focusing on the system's localized dynamic fluctuations, employing a block-analysis method similar to previous approaches used within the context of the four-point susceptibility ($\chi_4$) function~\cite{avila14,chakrabarty17}.
In ref.~\cite{PRE-RapidComm} we applied this method to two archetypal glass-forming systems:  computer simulations of the Kob-Andersen mixture~\cite{kob95a} and
confocal microscopy data of colloidal suspensions~\cite{weeks00}.
For  thermodynamic conditions for which motion is homogeneous in time and space (i.e. particle motion is not significantly correlated), we corroborated the expected behavior that the normalized dynamic fluctuations scale with a $N^{-1/2}$ power law decay. However as the relaxation enters the glassy regime,  the appearance of regions of correlated mobile particles makes the spatially localized dynamic fluctuations depart from such trivial behavior, decaying much slower with system size.
In this work, we apply the same methodology to computer simulations of liquid water.
By a careful study of the size-dependence of the molecular dynamic fluctuations, we  show the existence of an initial  power law decay  (that  gets progressively slower as we supercool the system) before the trivial system-size dependence is recovered at large $N$. The
cross-over to the   $N^{-1/2}$  regime  provides   an upper limit to the size of the largest spatially correlated relaxing regions. Additionally, we  demonstrate that  this regime is approached at larger $N$ values  as temperature $T$ decreases, suggesting a clear increase in the length scale of spatial heterogeneity on supercooling.

\section{Simulation details}

We perform NVT simulations using the TIP4P/2005~\cite{abascal05-jcp} model of water, which has emerged as the present-day optimal rigid water model~\cite{vega11-pccp}.
All simulations are conducted utilizing GROMACS 5.1.4~\cite{vanderspoel05-jcc} 
with a velocitiy-verlet integrator using a timestep of 1~fs.
The temperature is controlled using a  Nos\'{e}-Hoover thermostat~\cite{nose84-mp,hoover85-pra} while the coulombic interactions is evaluated  using a particle mesh Ewald treatment~\cite{essmann95-jcp} with a Fourier spacing of 0.1~nm.
The bond constraints are maintained using the LINCS (Linear Constraint Solver) algorithm~\cite{hess08-jctp}.
For both the Lennard-Jones and the real space Coulomb interactions an identical cut-off $r_\text{cut}=0.9$~nm is used.
Lennard-Jones interactions beyond $r_\text{cut}$ have been included assuming a uniform fluid density.
The TIP4P/2005 system consists of $N_{max}=250000$ molecules in a cubic box at density 0.95~g/cm$^3$ and it was studied at several $T$,
ranging from 230 to 360~K.  We have chosen to investigate the $\rho=0.95$ ~g/cm$^3$ isochore to avoid interference of the dynamics from the possible presence  of a liquid-liquid critical point, predicted to be above the $\rho=1.00$ ~g/cm$^3$ isochore~\cite{abascal10-jcp,sumi13,singh16,biddle17,handle2018potential}.

\section{Results}

The starting point of the method  is the observation of dynamic intermittency in molecular motion ~\cite{PRE-RapidComm}. Following prior work~\cite{ohmine1993fluctuation,appignanesi06,appignanesi09,water_PRE_F, appignanesi06b,appignanesi11,appignanesi07,PRE-RapidComm},
we compute a distance matrix $\Delta^2_{\cal S}(t',t'')$,
which represents the average of the squared molecular
displacements between times $t'$ and $t''$ of a collection of
$N$ water molecules belonging to a predefined set ${\cal S}$ (${\cal S}$ may be the
entire system or a subsystem, a  subvolume of the simulated system):
\begin{eqnarray}
\Delta^2_{\cal S}(t',t'') &\equiv & \frac{1}{N} \sum_{i=1}^{N}
| \vec{r}_i(t') - \vec{r}_i(t'')|^2_{i \in {\cal S}}\\
&=& \langle | \vec{r}_i(t') - \vec{r}_i(t'')|^2 \rangle_{i \in {\cal S}}
\end{eqnarray}
where the angle brackets indicate an average over the $N$ molecules
in $ {\cal S}$.  Further averaging $\Delta^2_{\cal S}(t',t'')$ over all pairs $t'$
and $t''$ such that $t''-t'=\Delta t$ yields the well-known
average mean square displacement $MSD(\Delta t)$ of the molecules in
$ {\cal S}$. More precisely, $MSD(\Delta t) = \langle \Delta^2_{\cal S}(t',t'')
\rangle_{t''-t'=\Delta t}$, where the average is over $t',t''$
with fixed time interval $\Delta t=|t''-t'|$ and also over all of
the particles in $ {\cal S}$.  Under stationary dynamics and for a sufficiently large $\Delta t$,
$\lim_{\Delta t \rightarrow \infty}\Delta^2_{\cal S}(t',t'+\Delta t) =
MSD(\Delta t)$.

\begin{figure}[th]
				{\includegraphics[width=9.5cm]{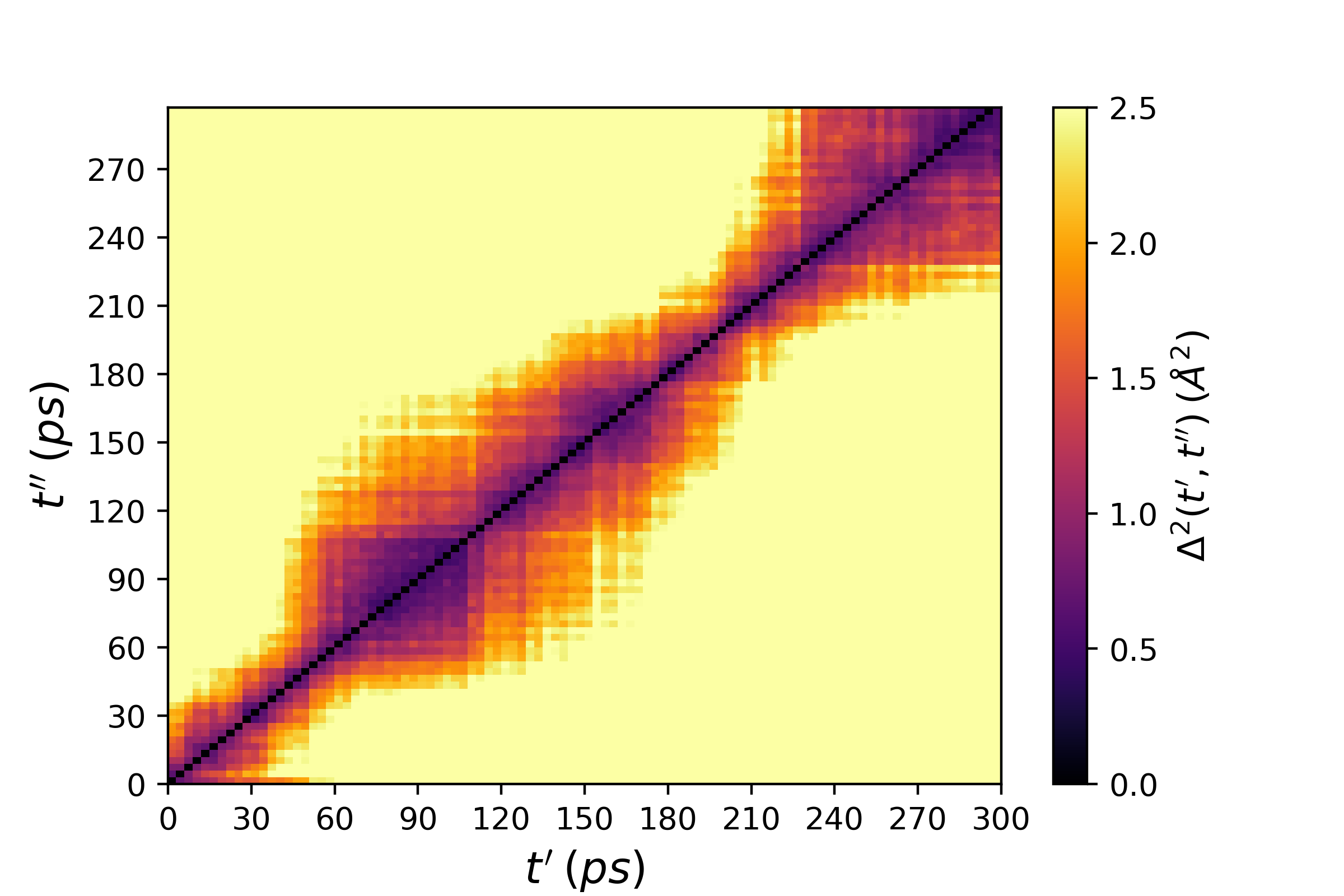}}
				{\includegraphics[width=9.5cm]{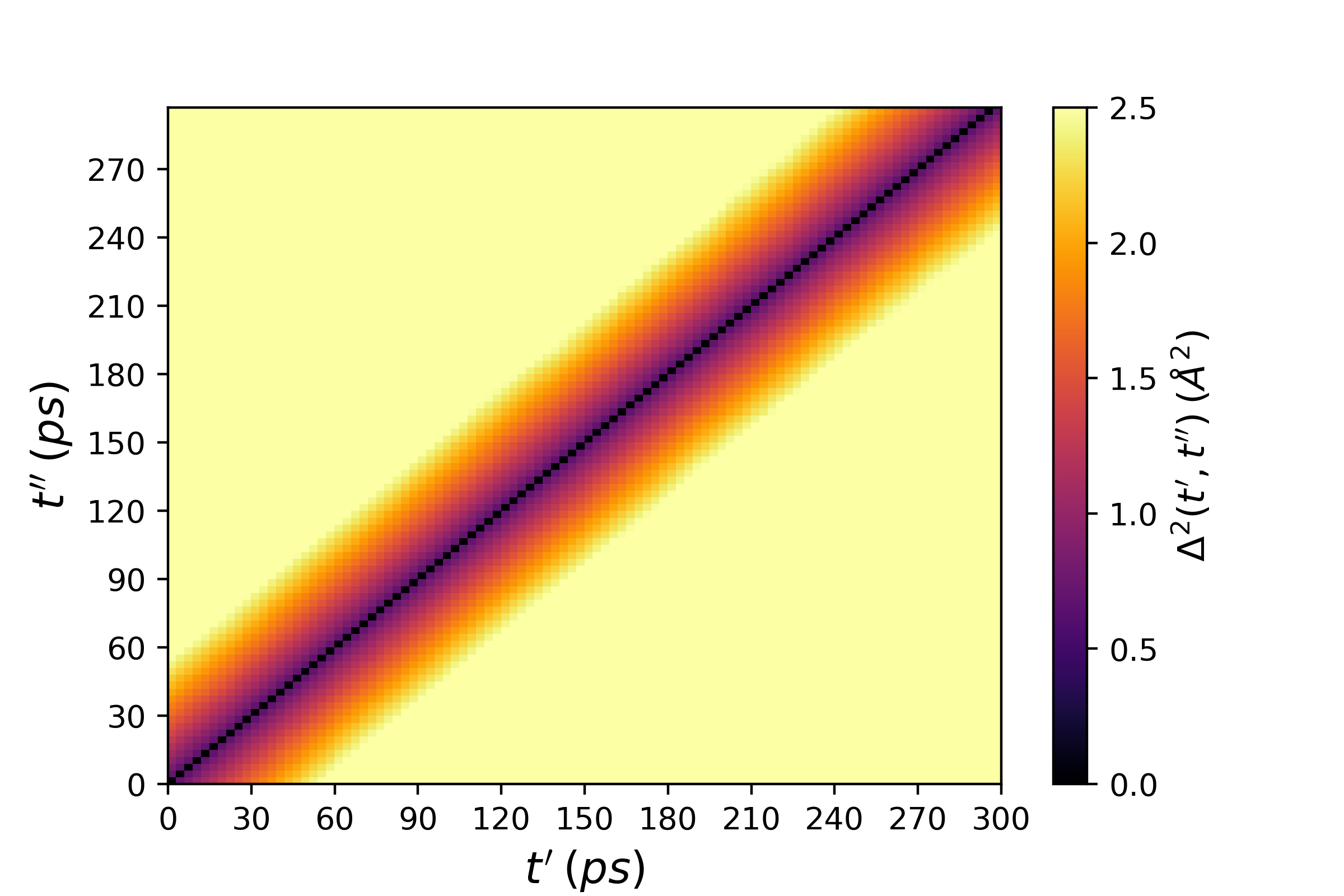}}
\caption{(a) Contour plot of the distance matrix
$\Delta^2_{\cal S}(t',t'')$ for a $TIP4P/2005$ water cubic  subsystem  ${\cal S}$ 
represented by a compact cubic block containing $N=250$  molecules
extracted from the  simulated $N_{\rm max}=250000$  system).
(b) Contour plot of the
same system for the full $N_{\rm max}=250000$ molecules simulation.
}
\label{dm}
\end{figure}

For small systems under glassy relaxation conditions, $\Delta_{\cal S}^2$
has temporal fluctuations, as shown in Fig.~\ref{dm}(a) for
a subsystem  ${\cal S}$ corresponding to  $N=250$ water molecules at $T=230K$. Darker regions indicate the existence of time intervals $(t',t'')$
over which this subsystem has relatively little particle
motion to then undergo rapid bursts of mobility. The latter events have been shown to involve the correlated large displacement of a relatively compact cluster of molecules that drive the system from one metabasin of its potential energy surface to a neighboring one ~\cite{appignanesi06,appignanesi09,water_PRE_F, appignanesi06b,appignanesi11,appignanesi07,PRE-RapidComm}. Since different regions within a large sample would suffer these relaxing events at different times, the island structure of the distance matrix  begins to be washed out as we increase the size  of the subsystem under study ~\cite{appignanesi09} (the spatial fluctuations average out such that
$\lim_{N \rightarrow \infty}\Delta^2_{\cal S}(t',t'+\Delta t)
= MSD(\Delta t))$. In other words, on increasing the subsystem 
size well beyond any dynamic correlation length, the independent behavior of the different regions of the system that are located  sufficiently far apart ~\cite{la2004static} make $\Delta^2_{\cal S}$ to appear much smoother at any given time,
as shown in Fig.~\ref{dm}(b) for the entire system. In turn, as $T$ is decreased, it is expected that the sizes of the correlated relaxing regions increase and, thus, we have to go to larger subsystem sizes in order to get a smooth distance matrix.

In a similar fashion as done for studies based on block-analysis of the four-point susceptibility ($\chi_4$)
~\cite{avila14,chakrabarty17}, we focus on the way in which the large system limit is
reached, and how this relates to the spatial scale of dynamical
heterogeneities ~\cite{PRE-RapidComm}. Since the obvious features of Fig.~\ref{dm}(a) are the
large fluctuations that differentiate it from Fig.~\ref{dm}(b), we consider the normalized difference between
$\Delta^2_{\cal S}$ and the expectation for a large system ~\cite{PRE-RapidComm}, defined by: 
\begin{equation}
\Omega_{\cal S}^2(t',t'') = 
\frac{ [\Delta^2_{\cal S}(t',t'') -  MSD(\Delta t)]^2} {[ MSD(\Delta t)]^2}
\end{equation}
with the convention $\Delta t = |t'' - t'|$.  $\Omega_{\cal S}^2$
represents the matrix of normalized squared deviations from
the mean value for the squared displacements of the water molecules and will
be equal to zero when $\Delta^2_{\cal S}$ is calculated for sufficiently
large systems, for which time averages and space averages are
equivalent and $\Delta^2_{\cal S} = MSD$.  Otherwise, $\Omega_{\cal S}^2 > 0$
and larger values indicate larger deviations between $\Delta^2$
(local in both space and time) and the expectation for a large system
(that is, $MSD$, a quantity averaged over all space and all time). Thus, $\Omega^2_{\cal S}(t',t'')$ provides us with a measure of dynamic intermittency. In simple terms, it reflects how different is the distance matrix for a small subsystem of size $N$ at a given time (like that in Fig.~\ref{dm}(a)) from the the situation when the results are averaged in size (large system) or, equivalently, in time (an outcome consistent with that shown in Fig.~\ref{dm}(b)). In practice, for each subsystem of interest we calculate $\Omega^2_{\cal S}(t',t'')$ for all the matrix elements of its distance matrix (that, for a small subsystem would look like Fig.~\ref{dm}(a)). For time intervals when we are within an island (as the ones depicted in  Fig.~\ref{dm}(a)), the relaxation is virtually stuck and, thus, the measure reflects the deviation of the relaxation behavior of the subsystem from the corresponding expectation value for the large system (that is, $MSD$, the mean squared displacement value corresponding to such time interval). In turn, when we focus on time intervals framing an island transition, as that depicted in  Fig.~\ref{dm}(a), we are faced with a large burst of mobility that also deviates from the most modest value corresponding to $MSD$ for such time interval. Thus,  in the calculation of the $\Omega^2_{\cal S}$ function we compute squared deviations in order to sum up both the excess and defect contributions that originate from all time intervals or matrix elements. Additionally, we make the calculation relative to $MSD$ in order to be left with normalized dynamic fluctuations. 

It is noteworthy that the $\Omega^2_{\cal S}$ function is local both in space and time.  To focus
on the spatial dependence of the fluctuations, we need to 
integrate out the time dependence.   To do so we calculate the ratio of
the dispersion to the average~\cite{chandlerbook} for the molecular
squared displacements, partitioning the large system of $N_{\rm max}=250000$ molecules into distinct cubical boxes (blocks) containing $N$ molecules each and evaluating
the sum of $\Omega_{\cal S}^2(t',t'')$ over all time pairs ($t'$, $t''$) divided by the number of such
pairs for each of the boxes.   We then average the resulting number
over all boxes and finally take the square root of the result.  Repeating this
procedure for several $N$ values, 
yields the desired time-independent  quantity $\Omega(N)$.

As noted in a prior work ~\cite{PRE-RapidComm}, the magnitude of $\Omega(N)$ depends on the total time
studied, that is, the maximum of $|t'' - t'|$ that is included in the calculation. Large $|t''-t'|$ time intervals contribute with small values and, thus, make the function $\Omega(N)$ to decrease ~\cite{PRE-RapidComm}.  However, for a given data set, the magnitude of the function is not relevant since the  $N$-dependence is insensitive to the total time studied provided that such time is able to capture the temporal fluctuations present in the distance matrix ~\cite{PRE-RapidComm}. In other words, what matters is to include a few of the ``islands'' seen in Fig.~\ref{dm}(a) ~\cite{PRE-RapidComm}. Consistently, in this work we adopt a timescale that represents a good choice in order to render a satisfactory $\Omega(N)$ function for the data we have examined. We thus take a total time given by the time scale when, at each temperature, the $MSD$ equals the (squared) nearest neighbour distance (the first peak position in the O-O radial distribution function) that is, the time when all the water molecules in the system have on average moved one intermolecular distance. This value, that is not far from the timescale of the maximum in the time dependence of the non-gaussian parameter and of the $\alpha$-relaxation time, lies after the plateau of the mean squared displacement curve (just beyond the end of the caging regime), at the beginning of the diffusive regime. At such time all the molecules have been able, on average, to break their first neighbors confinement in order to perform a significant local relaxation event.

\begin{figure}[th]
	\centering
		{\includegraphics[width=8.0cm]{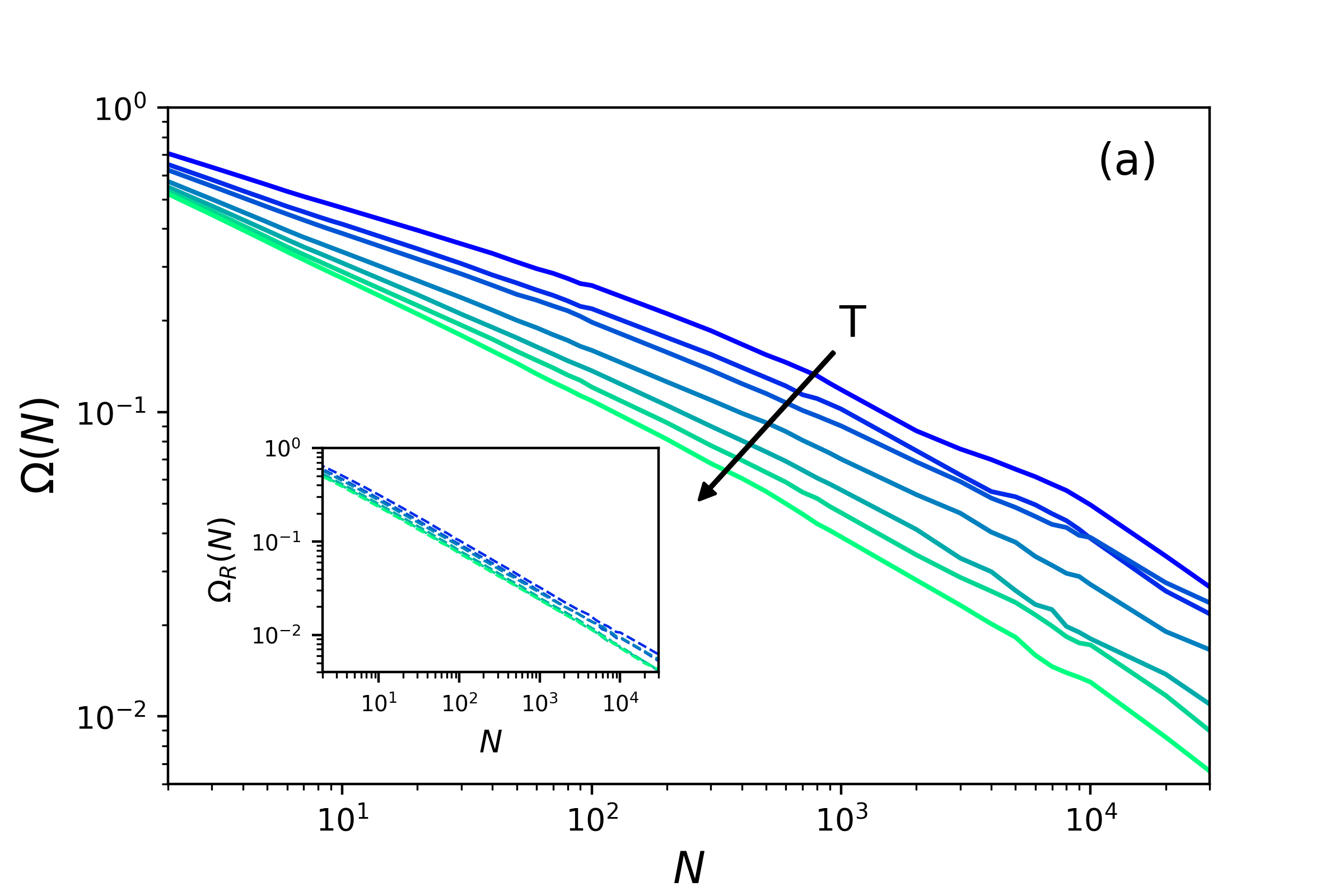}}
		{\includegraphics[width=8.5cm]{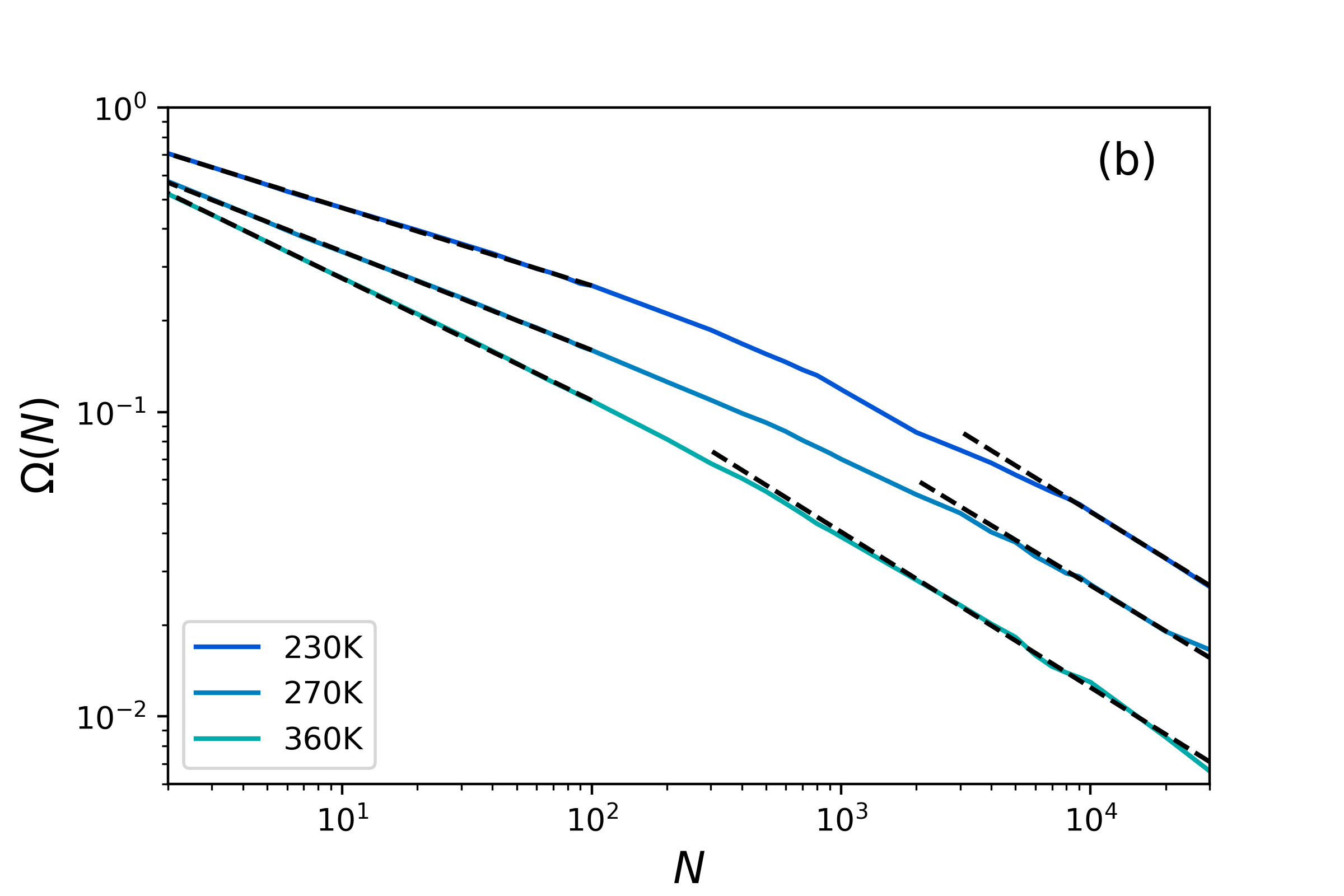}}
	\caption{(a) $\Omega({\rm N})$ and $\Omega_R({\rm N})$ (inset) as a function
	of subsystem size $N$ for  TIP4P/2005   for different temperatures as indicated.  In the case of $\Omega({\rm N})$ the 	$N$ molecules are part of the same compact subsystem (cubical block). In the case of
	$\Omega_R({\rm N})$ the $N$ particles are selected {\it randomly} among all the molecules in the system.  The size of the simulated system is always $N_{\rm max}=250000$. We show data up to roughly $N=N_{\rm max}/10$, as we need to average over at least 10 subsystems in order to get enough statistics to evaluate a reasonable $\Omega(N)$. (b) Same data as in (a) for $T$=230, 270 and 360~K displaying the two relevant regimes: The first one, with a decay that gets progressively slower as $T$ decreases and a last regime consistent with the trivial $N^{-1/2}$ size-dependence of the fluctuations. 
The latter regime is approached at higher $N$ as $T$ is lowered. 
	}
	\label{omega}
\end{figure}

Fig.~\ref{omega} displays the function $\Omega(N)$ for TIP4P/2005 water at temperatures $T$=230, 240, 250, 270, 300, 330, and 360~K. 
Similar to the simulations of the Kob-Andersen Lennard-Jones mixture and the experiments on colloidal suspensions we studied before~\cite{PRE-RapidComm}, the dynamical
fluctuations average out for large subsystem sizes.
On cooling, 
 larger and larger subsystems are required before the dynamical fluctuations are
averaged out. We also include in Fig.~\ref{omega} the function $\Omega_R(N)$ computed
using $N$
randomly chosen particles within the simulated system, 
destroying by construction any correlation in the motion of nearby particles 
 (the subscript $R$ stresses the random choice). 
Direct inspection of  Fig.~\ref{omega} shows that for these curves the heterogeneity is quickly averaged out with $N$, following a power law whose exponent does not depend on $T$.   

The functional form of
the decay of $\Omega(N)$ with $N$ provides the most relevant piece of information~\cite{PRE-RapidComm}. Fig.~\ref{omega} shows that the randomly distributed
dynamical fluctuations quantified by $\Omega_R(N)$ display a trivial system-size dependence, that is, they yield the
typical $N^{-1/2}$ decay at all temperatures. This reflects that particle motion
is nearly spatially uncorrelated within a subsystem and so the
average of $\Delta^2_{\cal S}(N)$ converges to the large-system limit
$MSD$ as $N^{-1/2}$. In turn, when $\Omega(N)$ is evaluated within compact subsystems of size $N$, we get a completely different picture. In a similar fashion as obtained for the Lennard-Jones mixture and for the experimental data on colloidal suspensions~\cite{PRE-RapidComm}, a clear departure from this trivial behavior is observed as temperature is decreased since the decay of
$\Omega(N)$ gets progressively slower. This significant enhancement of the spatially localized dynamical
fluctuations, persisting at large system
sizes, reflects the existence of regions of correlated mobile
particles, an effect that is more pronounced upon supercooling
\cite{glotzer00,kob97,donati98,donati99,weeks00,weeks07cor}. As temperature increases, we observe from Fig.~\ref{omega} that the spatially localized dynamic
fluctuations display a
size scaling dependence progressively closer to the usual $N^{-1/2}$ scaling law.


 At any given $T$, it is expected that the system presents a whole distribution of sizes of regions of correlated mobile particles. The sizes of such regions are, in turn, expected to increase with the degree of supercooling. As already discussed above, if we consider small subsystems within a large system, these regions of correlated mobile particles would govern the relaxation and, thus, the deviations from the large-system expectation value would be significant. However, as we focus on subsystems of progressively larger sizes, larger than the typical sizes of the regions of correlated mobile particles (that is, when the collectively relaxing regions are small as compared to the blocks), we expect that this behavior begins to be averaged out until the decay reverts to the trivial scaling down. A careful study of subsystems within a large total system would, thus, enable us to quantify the way in which such transition to the trivial regime occurs at larger subsystem size, $N$, as temperature is decreased. Thus, in Fig.~\ref{omega}-(b) we plot again the function $\Omega(N)$ (that is, for the block analysis) for temperatures $T=230, 270$ and $360$~K. From such figure it is immediately evident that the curves indeed present two clearly different regimes: a first power-law regime for the low $N$ region where the relaxation is dominated by the spatially localized dynamic fluctuations arising form the collective motions, while at large $N$ the curves revert to the trivial system-size scaling (power law exponent of ${-1/2}$). The latter regime is indeed approached at larger $N$ values as $T$ decreases. 

\begin{figure}[th]
				{\includegraphics[width=8.5cm]{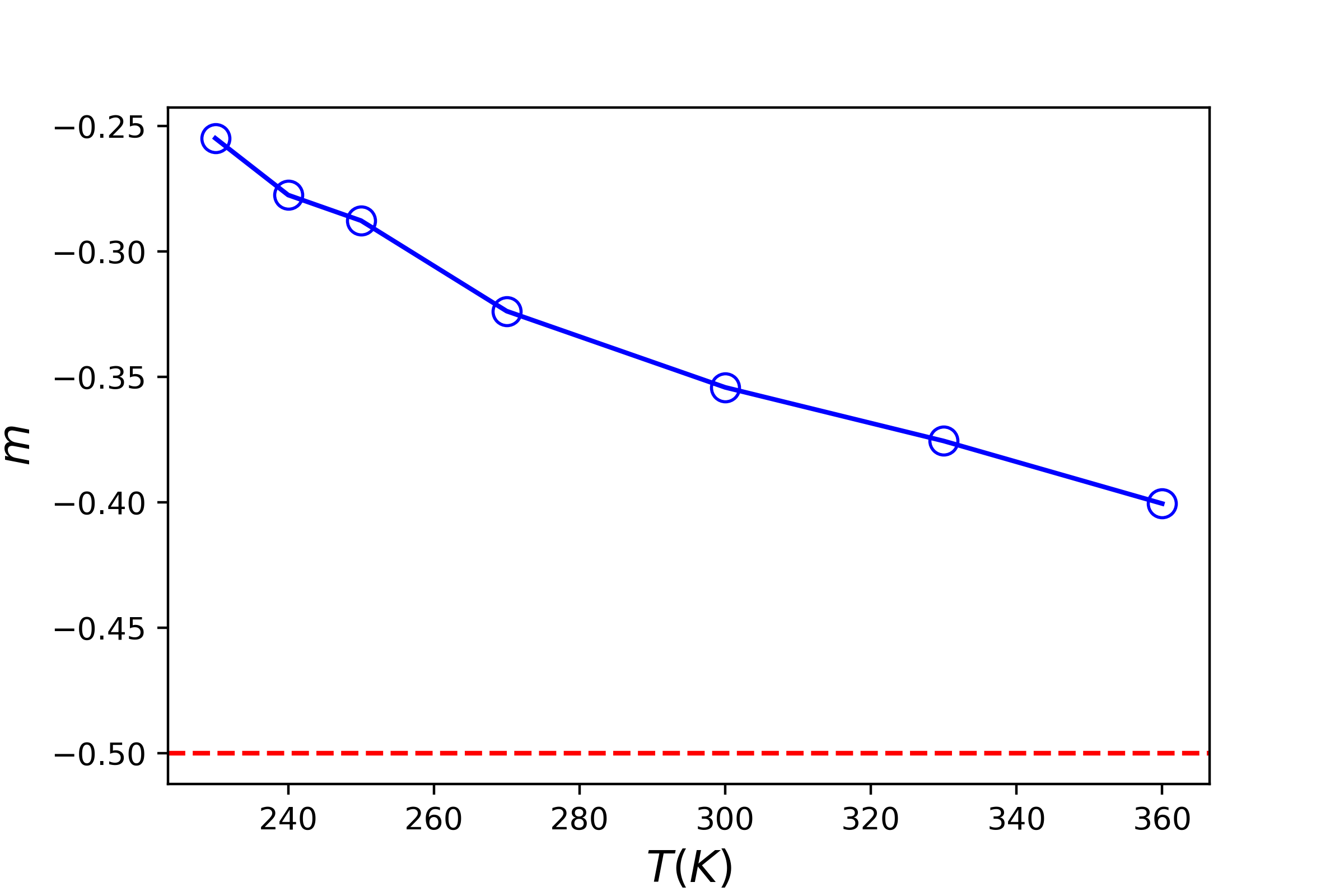}}
\caption{Decay exponent for the low $N$ regime of $\Omega(N)$ (slopes $m$ from the logarithmic plots of Fig.~\ref{omega} (a)) as a function of temperature. The dashed line indicates trivial system-size scaling.
}
\label{slopes}
\end{figure}

In  Fig.~\ref{slopes} we plot the decay exponent (defined as the slope $m$ from the logarithmic plot of Fig.~\ref{omega}-(a)) of the first (small $N$) regime of $\Omega(N)$ as a function of $T$. For the lowest $T$, $T=230$~K, $m \approx -0.25$ depicting the reluctance of the dynamic fluctuations to fall with increasing size. This value decreases towards the trivial decay ($m=-0.5$) as $T$ is incremented (for the largest $T$ studied, $T=360$~K, $m$ is around $-0.4$). 

\begin{figure}[th]
				{\includegraphics[width=8.5cm]{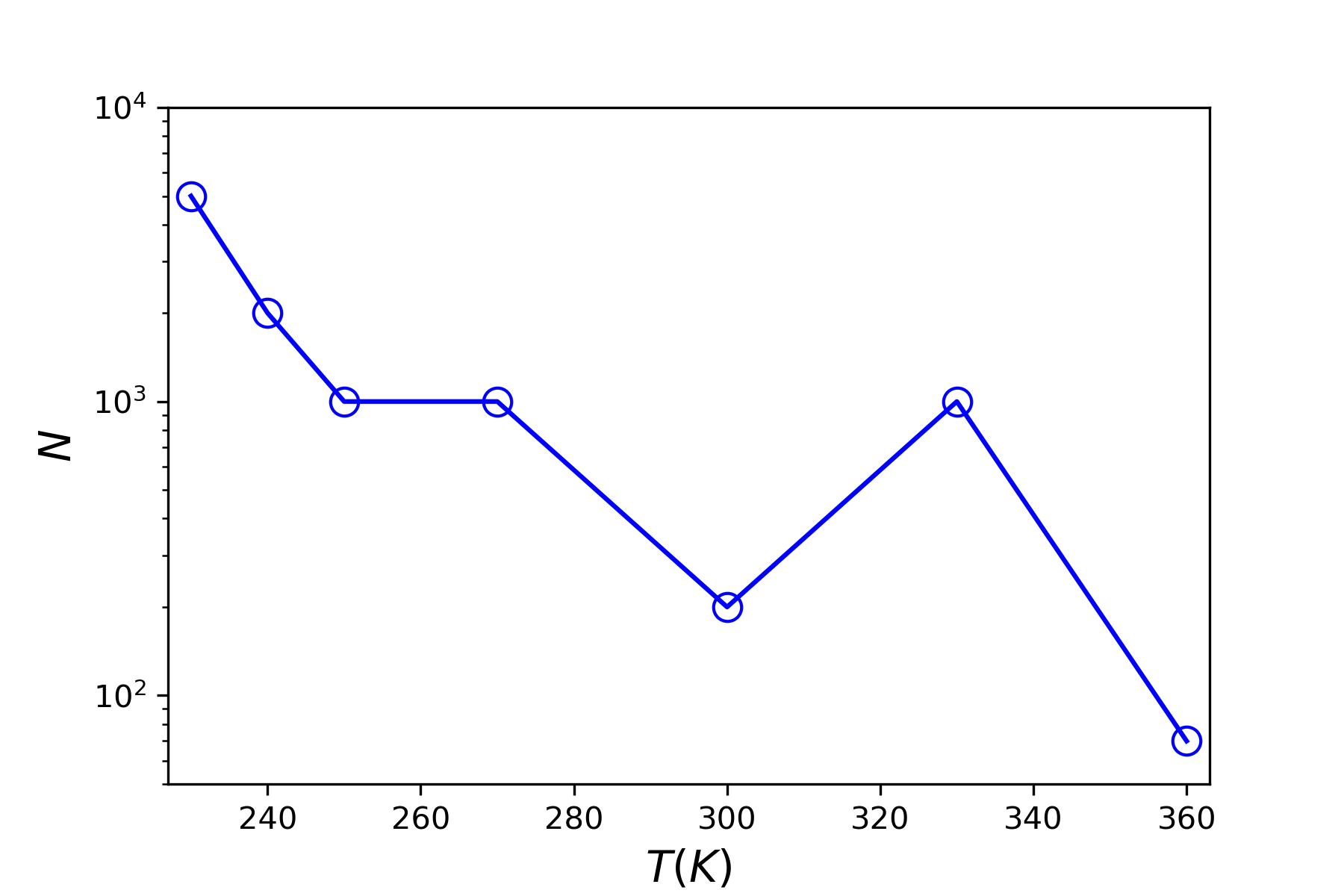}}
\caption{Estimation of the length scale of approaching to the trivial $N^{-1/2}$ scaling regime for the different temperatures studied.}
\label{3sigma}
\end{figure}

In turn, as already indicated, the  $N$ value where $\Omega(N)$ crosses to the trivial  $N^{-1/2}$ decay indicates that the large-limit behavior has been reached. Such crossover implies that the subsystem is now composed by a sufficiently large number of independently relaxing regions and, thus, represents the length scale at which the influence of the collective relaxation regions is averaged out.
Qualitatively, direct inspection of Fig.~\ref{omega} (b) makes it clear that this happens at a much higher $N$ as temperature is lowered. To quantitatively estimate this length scale we now study in detail the large $N$ decay of $\Omega(N)$ (to avoid possible statistical errors we consider the $\Omega(N)$ function at up to $N=N_{\rm max}/10$ to get  at least 10 subsystems to evaluate a reasonable value). Starting at  $N=N_{\rm max}/10$, we extend the theoretical decay regime to lower $N$ values by imposing a $-0.5$ exponent (that is, a $-0.5$ slope in the logarithmic plot of $\Omega(N)$ vs $N$), provided the correlation coefficient is larger than $0.99$. We then calculate the value of $N$ for which $\Omega(N)$ deviates more than $3 \sigma$ from this behavior, which marks the point of departure from the  trivial regime as $N$ is decreased.  Fig.~\ref{3sigma} displays the results. The approach to the trivial size decay of the fluctuations and, thus, the length scale of maximal influence of the regions of correlated collective relaxation indeed depends strongly on temperature. Specifically, we find that for the lowest temperature studied, $T=230$~K, it occurs at around $N=5000$, a size almost two orders of magnitude larger than the situation for $T=360$~K.

\section{Conclusions}

In this work we have applied a measure of spatial and
temporal dynamic heterogeneity to liquid and supercooled water by studying the evolution of dynamic fluctuations averaged over different space
lengths and time scales. We have corroborated previous results in other glassy systems indicating that the appearance, upon supercooling, of regions of correlated mobile molecules make the system present significant spatially localized dynamic fluctuations. A careful study of the size dependence of such dynamic fluctuations has now enabled us to distinguish two clearly different regimes: An initial regime in which fluctuations decay unusually slowly with system size, a behavior that is more conspicuous as temperature is decreased while, at large length scales, the behavior recovers the trivial scaling down of dynamic fluctuations characterized by the typical $N^{-1/2}$ power law decay. The system size at which this final regime is approached  
significantly grows as $T$ decreases, reaching  values 100 times larger (in $N$) 
than the high-$T$ limit.  At  lowest temperatures studied, 
averaging out  the influence of the regions of correlated mobile molecules  requires approximatively thousand particles.

\begin{acknowledgments}
GAA SRA and JMMO acknowledge support form CONICET, UNS and
ANPCyT(PICT2015/1893). PHH acknowledges support from the Austrian Science Fund (FWF Erwin Schr\"{o}dinger Fellowship J3811 N34). 
\end{acknowledgments}


%
\end{document}